# Horizontal Test for BEPCII 500MHz Spare Cavity [*]

This article has been contributed to **Chinese Physics C**

MI Zhenghui (米正辉)[1;2;1)] SUN Yi(孙毅)[2] WANG Guangwei(王光伟)[2] PAN Weimin(潘卫民)[2]
Li Zhongquan(李中泉)[2] DAI Jianping(戴建枰)[2] MA Qiang(马强)[2] LIN Haiying(林海英)[2]
XU bo(徐波)[2] HUANG Hong(黄泓)[2] WANG Qunyao(王群要)[2] XU Yufen(许玉芬)[2]
ZHAO Guangyuan(赵光远)[2] HUANG Tongming(黄彤明)[2] SHA Peng(沙鹏)[2]
LIU Yaping(刘亚萍)[2] MENG Fanbo(孟繁博)[1;2] QIU Feng(邱丰)[1;2] LI Han(李菡)[1;2]
CHEN Xu(陈旭)[1;2] ZHAO Danyang(赵丹阳)[1;2] ZHANG Juan(张娟)[1;2]

1 (Graduate University of the Chinese Academy of Sciences, Beijing 100049, China)
2 (Institute of High Energy Physics, CAS, Beijing 100049, China)

**Abstract**：The horizontal test for BEPCII 500MHz spare cavity has been completed in IHEP. The max voltage of the cavity is 2.17MV, $Q_0$ is $5.78\times10^8$. This paper mainly reports coupler aging parameters before the horizontal test, as well as the parameters during cool-down the change of frequency、loaded Q、vacuum and the deformation of the cavity with temperature. And the test results were analyzed.

## 1. Introduction

A 500MHz superconducting cavity was fabricated at IHEP in 2011. After completion of assembly, a horizontal test is carried out for the spare cavity in IHEP. The spare superconducting cavity is the first 500MHz spare superconducting cavity of BEPCII. All key parts of the cavity were made in china. The performances of the spare cavity exceeded the design values. The max voltage of the cavity is up to 2.17MV.

## 2. Structure of spare cavity

The BEPCII spare superconducting cavity consists of niobium cavity, cryostat, high power input coupler, tuner and high-order mode absorber. The spare superconducting cavity structure is shown in figure 1. Superconducting cavity assembly was done by the Institute of high energy physics researchers independently.[1]

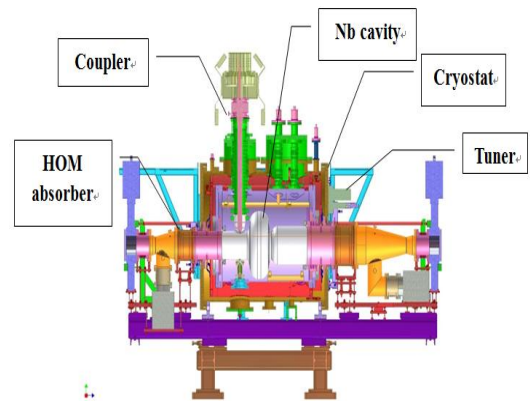

Figure 1: The structure of cavity

## 3. Test stages

### Room temperature

After completing the assembly of the cavity, the cavity was transported to the horizontal test room, where the high

---

[*] Supported by the "Facility repair project" of CAS, under Grant No. K71G710Y21
1) E-mail: mizh@ihep.ac.cn

power horizontal test was done. The horizontal test is an overall test of the cavity without a beam. The process of the test is as follows:

Check all the interlocks to confirm that all the signal equipment works well. The interlocks are shown in table 1.

Table 1: Inter locks

| Fast lock | Normal lock |
|---|---|
| Coupler Vacuum | Cooling Water |
| Cavity Vacuum | Air Flow |
| Arc | Gas Flow |
| Quench | He Level |
| P+ | He Pressure |
|  | Temperature |

Tuner test: At the room temperature, the tuner performance is tested. It is the most important test at the room temperature. The tuner of the spare cavity is composed of a main motor and a piezoelectric oscillator. The main motor has a large stroke, while the operation is slow. On the other hand, the piezoelectric oscillator has a small stroke and a fast operation. If there are something wrong with the tuner, it is found at this stage. For the tuner of BEPCII spare cavity, it was found that the movement was smooth.[1]

Coupler test and aging: Operation of the tuner was performed in order to detune the cavity, then aging of the coupler was performed. The results of the coupler aging is shown in 2(a). This figure shows the aging of coupler from October 1st to 3rd. The max input power exceeded 106KW, which is also the total reflected. The coupler was preheated before aging to avoid cracking of the coupler window. Both pulsed and continuous waves were applied repeatedly as this method can save aging time and improve safety performance. The first aging condition of the cavity vacuum is $2.2\times10^{-6}$pa, and the coupler vacuum is $4.74\times10^{-6}$pa; The second aging condition of the cavity vacuum is $8.2\times10^{-7}$pa, and the coupler vacuum is $1.64\times10^{-6}$pa; The third aging condition of the cavity vacuum is $7.1\times10^{-7}$pa, and the coupler vacuum is $1.18\times10^{-6}$pa.

During test, the number of interlocks caused by Arc、coupler vacuum、cavity vacuum is shown in figure 2(b). From figure 2(b) we know when aging the coupler at low input power, interlock protection almost caused by cavity

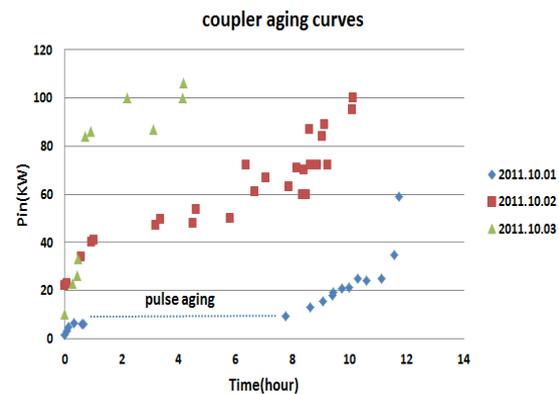

(a)

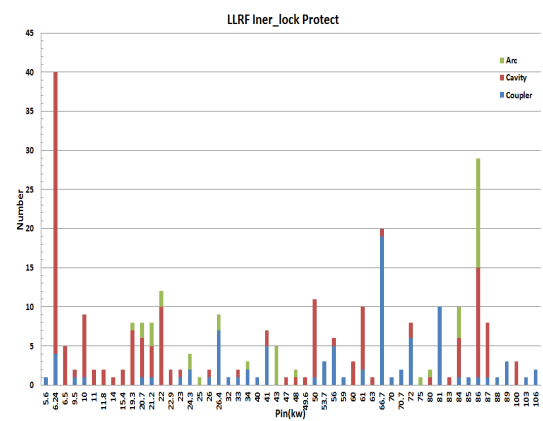

(b)

Figure 2: (a) coupler aging curves

(b) the number of interlock protect

vacuum. In high input power, interlock protection were almost caused

by coupler vacuum. Arc mostly happened between 19kw and 43kw, and between 75kw and 87kw.[2]

**During cool-down**

During cool-down, the frequency and the loaded Q value of the spare cavity were monitored with a network analyzer every few hours. Figure 3(a) shows the result of the frequency and loaded Q change during cool-down.

Figure 3(a) shows during cool-down the frequency of spare cavity increases. The loaded Q of the cavity also increases. After the cavity temperature reached 80K, the loaded Q increased quickly. At 4.4K the loaded Q of the spare cavity is $1.3 \times 10^5$, and the frequency is 499.507MHz.

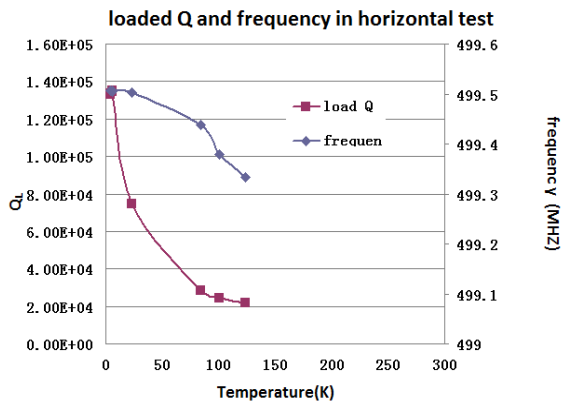

(a)

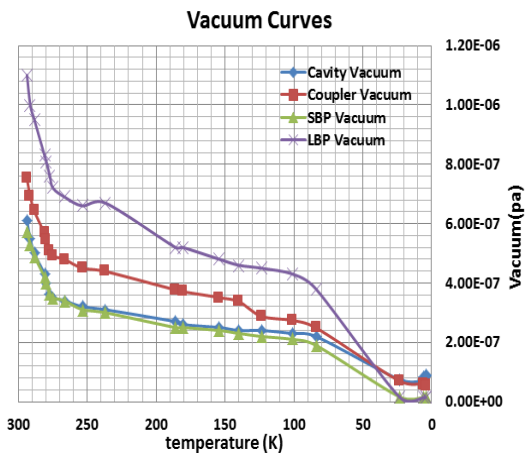

(b)

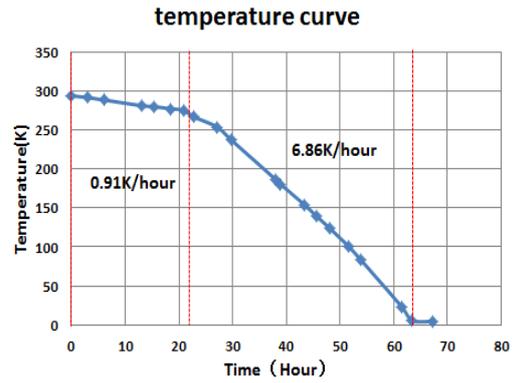

(c)

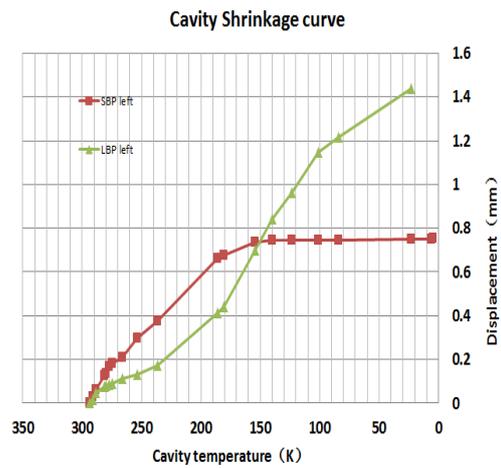

(d)

Figure 3: (a)The frequency and loaded Q; (b) The change of vacuum during cool-down;(c) The time profile of cavity temperature during cool-down; (d)shrinkage of spare cavity

Figure 3(b) shows that the vacuum of the cavity、coupler、SBP、LBP are all very well. There was no evidence of vacuum leakage during cool-down. From figure 3(c) shows the time profile of the cavity temperature during cool-down. In the first 20 hours the cool-down rate was about 0.91K/hour, which was intentionally very slow to avoid vacuum leakage. In the next 40 hours, the cool-down rate was about 6.8K/hour. [1]

In order to analyze the shrinkage situation of the spare cavity, two displacement meters were respectively installed on the SPB and LBP ends, before cool-down. As shown in figure 3(d), we can see that during the cavity temperature decrease from 300K to 215K, SBP end shrinkage was about 0.752mm. After that, the displacement did not change anymore. The LBP ends displacement changed, during the entire cool-down process, was about 1.438 mm, larger than SBP ends shrinkage.

### 4.4k

After the cavity temperature decreased to 4.4K, operation of the tuner demonstrated the resonant frequency of the cavity 499.8MHz. During the tuning process through network analyzer observed the change of cavity frequency with load as shown in figure 4. From figure 4 can see that the cavity frequency has a linear relation with load, and the cavity of the elastic coefficient is 1KHz /kg.

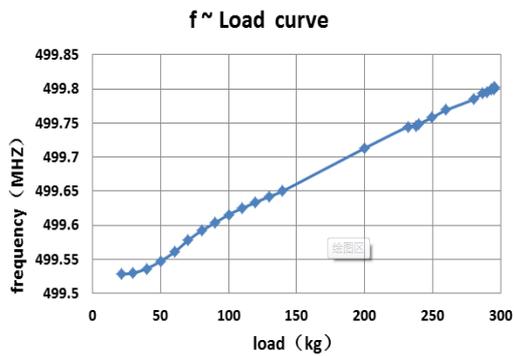

Figure 4: Frequency-load curve

When the pare cavity was tested at high power, the coupler vacuum deteriorated. Several hours later, the test was repeated, and the coupler vacuum returned to normal. This suggests that the coupler aging time was perhaps too short previously.[3]

Spare cavity $Q_0$ value has been calculated through the consumption of total calories. In order to guarantee the accurate measurement of $Q_0$ value, the cavity voltage Vc and cavity loss Pc must be accurately calibrated. Therefore, before testing on, two methods were used to calibrate cavity voltage Vc. The first method is based on cavity voltage and input power relations; the other is based on through precise measurement of cavity voltage sampling signal Pickup power $P_{15D}$ and vertical test calibration $Q_{15D}$.

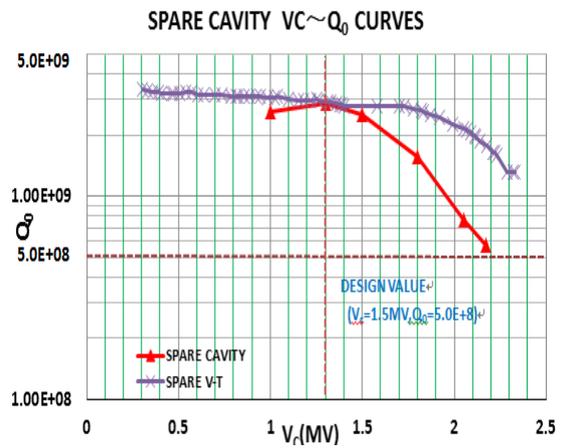

(a)

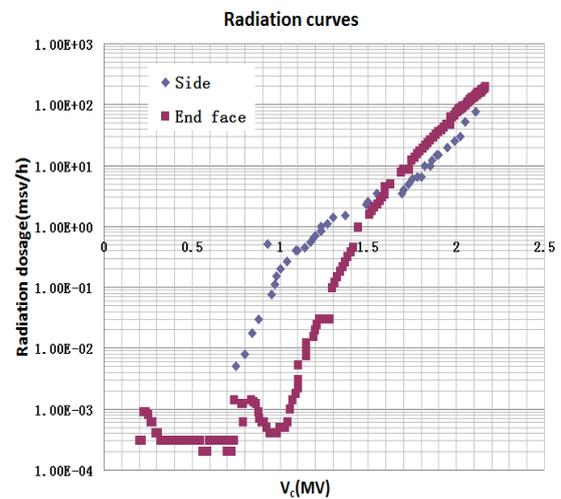

(b)

Figure 5: (a)The comparison of $Q_0$ vs. Vc curve between vertical and horizontal test.(b) radiation curves of cavity

Figure 5(a) shows the $Q_0$ value is lower in horizontal test than in vertical test. In horizontal test, $Q_0$ decreased rapidly when the cavity voltage was over 1.5MV, while in vertical test the $Q_0$ declined slowly when the cavity voltage was over 2MV.

Before testing, radiation instruments has been installed at SPB、LBP ends and side of cryostat. Figure 5(b) shows that the radiation is very low when spare cavity voltage is between 0 and 1MV. When the cavity voltage increased over 1MV, the radiation level increased quickly. The radiation level is higher in

Table 2: BEPCII Spare Cavity main parameters

| | |
|---|---|
| Vacc（Max） | 2.17MV |
| $Q_0$ | $7.67 \times 10^8$ (@2.0MV) |
| Work frequency | 499.8MHz |
| Frequency（@4.4K,free Load） | 499.507MHz |
| $Q_L$ | $1.33 \times 10^5$ |
| Static heat leakage | 41W |
| Coupler Power（Total reflection） | 100KW |
| Mechanical frequency | 46Hz |

Axial than in Transverse, with values of 197msv and 76msv respectively.

Table 3: BEPCII Spare Cavity main parameters

| | Voltage（MV） | $Q_0$ |
|---|---|---|
| Spare cavity | 2.05 | $7.67 \times 10^8$ |
| East cavity | 2.0 | $5.45 \times 10^8$ |
| West cavity | 2.05 | $9.5 \times 10^8$ |

As shown in table 3, spare cavity performance is superior to the East cavity, but lower than west cavity.

## 4. Conclusion

1. BEPCII 500MHZ $Q_0$ value is 5.78E+8 at 2.17MV, exceeding the design requirement that cavity $Q_0$ value not be less than the requirements of 5.0E+8 at 1.5MV.

2. The spare cavity $Q_0$ in the horizontal test was lower than that in the vertical test. When cavity voltage was over 1.5MV, $Q_0$ decreased greatly, and the radiation level rose substantially. These results suggest that pollution may have entered the superconducting cavity compromising the inner surface of the cavity during the cavity assembly process. The cavity performance would be further improved with aging time[4].

3. BEPCII spare superconducting cavity's frequency is 499.507MHz at 4.4K free load condition. When load is 295kg, the frequency of the spare cavity is 499.8014MHz, without compensating springs. Superconducting cavity processing, polishing, and assembly would impact the cavity frequency, but at all stages we could control the frequency in a Proper value. This shows that in the manufacturing process of the superconducting cavity, we have had bigger breakthrough.

## 5. Acknowledgement

Thanks to the test group. Special thanks are given to KEK Three experts Shinji MITSUNOBU,Takaaki FURUYA, KAZUNORI AKAI, thanks them for the help of BEPCII spare superconducting cavity and support for work over the years.